\documentclass[twocolumn,showpacs,preprintnumbers,amsmath,amssymb]{revtex4}

\usepackage{graphicx}
\usepackage{dcolumn}
\usepackage{bm}

\def\FigFactor{0.47}
\def\sqrtsNN{\mbox{$\sqrt{s_\mathrm{NN}}$}}

\begin{document}

\preprint{LBNL-54700 Conf.}

\title{Anisotropic flow in the forward directions at \boldmath$\sqrt{s_\mathrm{NN}} = 200$\unboldmath\,GeV}

\author{Markus\ D.\ Oldenburg}
\email{MDOldenburg@lbl.gov}
\affiliation{%
(for the STAR Collaboration)\\
\\
Lawrence Berkeley National Laboratory\\
Nuclear Science Division\\
MS70R0319, One Cyclotron Road\\
Berkeley, California, 94720, USA}%

\date{March 17, 2004}

\begin{abstract}
The addition of the two Forward TPCs to the STAR detector allows one to measure
anisotropic flow at forward pseudorapidities. This made possible the first
measurement of directed flow at collision energies of $\sqrtsNN =
200$\,GeV. PHOBOS' results on elliptic flow at forward rapidities were
confirmed, and the sign of $v_2$ was determined to be positive for the first
time at RHIC energies. The higher harmonic, $v_4$, is consistent with the
recently suggested $v_2^2$ scaling behavior.

This write-up contains results presented as a poster \cite{poster} at the Quark
Matter conference in Oakland, California in January 2004.
\end{abstract}


\maketitle

\section{\label{intro}Introduction}

In non-central heavy-ion collisions the initial spatial anisotropy of the
collision region translates into a final state anisotropy in momentum space. In
a hydrodynamical picture this is believed to be due to pressure gradients in the
dense medium which lead to collective motion --- so called transverse flow ---
of the generated particles.

The simplest way of characterizing these final state anisotropies is to perform
a Fourier decomposition on the particle's emission angles $\phi$ with respect to
the reaction plane $\Psi_{RP}$~\cite{vol}. The reaction plane is given by the
incident beam direction and the impact parameter and it is experimentally not
known \emph{a priori}. It has to be estimated for every event by looking at the
anisotropy of particle emission itself~\cite{fourier}. This leads to a finite
resolution of the measured event plane which one has to correct for.

Spurious contributions to the measured transverse flow signal are particle
correlations due to non-flow effects (e.\,g.\ resonance decays). To cope with
these, several new methods of the anisotopic flow analysis, based either on
cumulants~\cite{BDO1,BDO2} or on Lee-Yang zeros~\cite{LeeYangZeros}, have been
proposed.

\section{\label{expsetup}Experimental setup}

The two Forward TPCs (FTPCs~\cite{Ftpc}) of the STAR experiment~\cite{STAR}
extend the pseudorapidity coverage of STAR into the region $2.5<|\eta|<4.0$. The
pseudorapidity resolution of these radial drift chambers is better than 5\% for
their full acceptance. During RHIC run 2 about 70~thousand Au+Au collisions at a
center of mass energy of $\sqrtsNN = 200$\,GeV were taken with both FTPCs and
the STAR TPC~\cite{Tpc}.

\section{Measurements}

\subsection{Directed Flow \boldmath$v_1$\unboldmath}
\begin{figure}[ht]
\resizebox{0.5\textwidth}{!}{\includegraphics{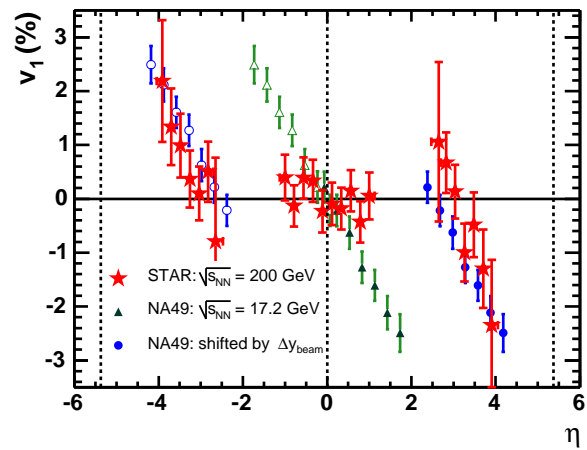}}
\caption{\label{v1}
Directed flow measured at $\sqrtsNN = 200$\,GeV compared to results measured by
NA49 in fixed-target collisions at $158A$\,GeV. The red stars show our
measurement of $v_1\{3\}$. We see a clear signal of directed flow at forward
pseudorapidities. While our signal differs greatly from the measured result by
NA49~\cite{NA49Paper} (green triangles), there is a good agreement between the
two signals in the projectile frame.}
\end{figure}

The first measurement of directed flow at RHIC energies was recently
published~\cite{v1v4Paper} (see Fig.~\ref{v1}). It showed that while $v_1(\eta)$
is close to zero at mid-rapidities, the signal rises to a couple of percent near
pseudorapidity $|\eta| \approx 4$.

It was noted that our measurement greatly differs from the NA49
results~\cite{NA49Paper} at lower beam energies of $158A$\,GeV. But if the NA49
data are shifted and both measurements are seen in the projectile frame, they
look similar.

\subsection{Elliptic flow \boldmath$v_2$\unboldmath}
\begin{figure}[ht]
\resizebox{\FigFactor\textwidth}{!}{\includegraphics{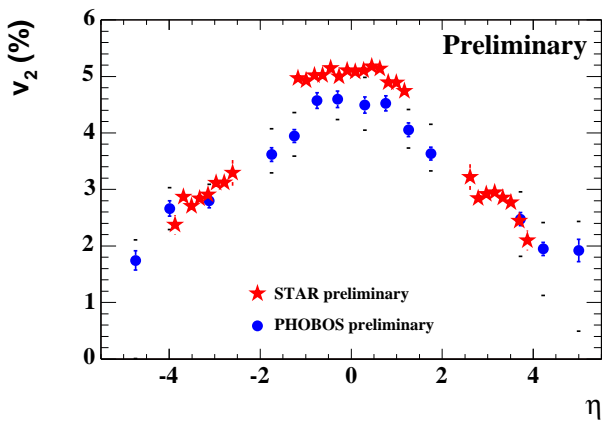}}
\caption{\label{v2phobos}
Elliptic flow $v_{2}$ vs.\ pseudorapidity $\eta$ at $\sqrtsNN = 200$\,GeV. The
measured $v_{2}(\eta)$ from the STAR collaboration (red stars) agrees very well
with the published~\cite{Phobosv2} (but preliminary) results by PHOBOS at
forward pseudorapidities: the fall-off from $\eta = 0$ to $|\eta| = 3$ is
confirmed.}
\end{figure}

The comparison of our new measurement on elliptic flow $v_2(\eta)$ at forward
pseudorapidities confirms the published result~\cite{Phobosv2} obtained by the
PHOBOS collaboration (see Fig.~\ref{v2phobos}) at high $\eta$. We observe a
similar fall-off by a factor of 1.8 comparing $v_2(\eta = 0)$ with $v_2(\eta
=3)$. Both measurements were done using the event plane method.

If we compare our results for $v_2$ obtained with the method of two-particle
cumulants, $v_2\{2\}$, to the four-particle cumulants, $v_2\{4\}$, we observe
almost no difference in the FTPC region, while the two-particle cumulant
measurement gives a 15\% higher signal in the TPC. Since four-particle cumulants
are much less prone to non-flow contributions we conclude that non-flow effects
are less strong in the forward regions.

\subsection{A new method to measure directed flow}
From the above measurements it became clear that the STAR TPC sitting at
mid-rapidity has very good capabilities to measure elliptic flow, while the
Forward TPCs allow to measure directed flow (which appears to be close to zero
at mid-rapidities).

\subsubsection{$v_1\left\{\mathrm{EP}1,\mathrm{EP}2\right\}$}
In order to utilize the method of Fourier decomposition but to reduce non-flow
contributions at the same time, we measured $v_1$ with respect to the first and
second order reaction plane $\Psi_1$ and $\Psi_2$, where $\Psi_1$ was determined
in the FTPCs while $\Psi_2$ was measured in the TPC. Within the recently
proposed notation (see~\cite{v1v4Paper}) we denote this measurement as
$v_1\left\{\mathrm{EP}1,\mathrm{EP}2\right\}$.
\begin{eqnarray}
v_1\{\mathrm{EP}1,\mathrm{EP}2\} =\;\;\;\;\;\;\;\;\;\;\;\;\;\;\;\;\;\;\;\;\;\;\;\;\;\;\;\;\;\;\;\;\;\;\;\;\;\;\;\;\;\;\;\;\;\;\;\;\;\;\;\;\;\;\;\\
\frac{\left\langle\cos\left(\phi+\Psi_1^{\mathrm{FTPC}}-2\Psi_2^{\mathrm{TPC}}\right)\right\rangle}{\sqrt{\left\langle\cos\left(\Psi_1^{\mathrm{FTPC}_1}+\Psi_1^{\mathrm{FTPC}_2}-2\Psi_2^{\mathrm{TPC}}\right)\right\rangle\cdot \mathrm{Res}(\Psi_2^{\mathrm{TPC}})}}\nonumber\;.
\end{eqnarray}

\begin{figure}[ht]
\resizebox{\FigFactor\textwidth}{!}{\includegraphics{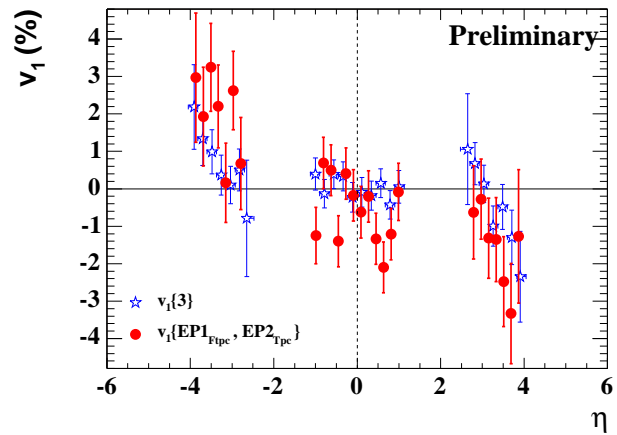}}
\caption{\label{v1Ep1Ep2}
Directed flow with respect to the first and second order reaction plane as a
function of pseudorapidity. The measurements of
$v_1\{\mathrm{EP}1,\mathrm{EP}2\}$ (red circles; centrality 20--60\%) agree
within the errors with the published results of $v_1\{3\}$ (centrality
10--70\%).}
\end{figure}

As shown in Fig.~\ref{v1Ep1Ep2}, the results are in reasonable agreement with
the published measurement obtained by the three-particle cumulant method
$v_1\{3\}$.

\subsubsection{The sign of $v_2$}
This new method provides an elegant tool to measure the sign of $v_2$, which was
assumed to be positive but had not yet been determined at RHIC energies. One of
the quantities involved in the measurement of $v_1\{\mathrm{EP}1,\mathrm{EP}2\}$
is approximately equal to the product of integrated values of $v_1^2$ and $v_2$:
\begin{eqnarray}
v_1^2\cdot v_2 \approx
\frac{\left\langle\cos\left(\Psi_1^{\mathrm{FTPC}_1}+\Psi_1^{\mathrm{FTPC}_2}-2\Psi_2^{\mathrm{TPC}}\right)\right\rangle}{\sqrt{M_{\mathrm{FTPC}_1}\cdot
M_{\mathrm{FTPC}_2}\cdot M_{\mathrm{TPC}}}}\;,
\end{eqnarray}
where $M_{\mathrm{FTPC}_1}$, $M_{\mathrm{FTPC}_2}$, and $M_{\mathrm{TPC}}$
denote the multiplicities for a given centrality bin in the two FTPCs and the
TPC, respectively.  Since $v_1^2$ is always positive, the sign of $v_1^2\cdot
v_2$ determines the sign of $v_2$.

\begin{figure}[ht]
\resizebox{0.5\textwidth}{!}{\includegraphics{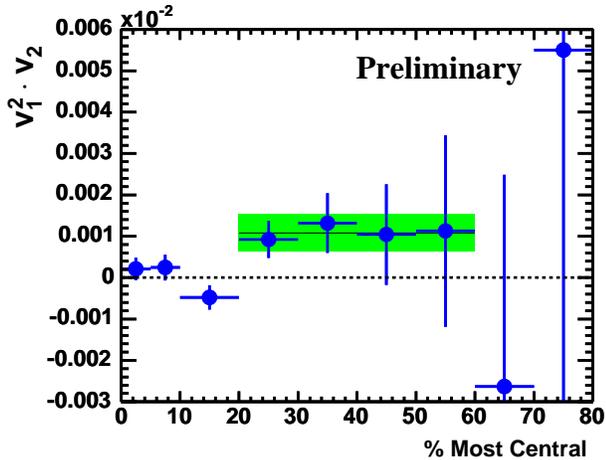}}
\caption{\label{v1v1v2}
The product of $v_1^2$ and $v_2$. Since the mean value of this quantity,
averaged over centralities 20--60\%, is $(1.08\pm0.46)\cdot 10^{-5}$, elliptic
flow is measured to be {\it in-plane}.}
\end{figure}

Averaged over centralities 20--60\% we measure $v_1^2\cdot v_2$ in
Fig.~\ref{v1v1v2} to be $(1.08\pm0.46) \cdot10^{-5}$. In this region the
expected non-flow contributions are much smaller than for the most central and
peripheral centrality bins. Therefore the sign of $v_2$ is determined to be
positive: {\it In-plane} elliptic flow is confirmed. (This stated value for
$v_1^2\cdot v_2$ and its uncertainty is based on an approximation that does not
affect the statistical significance of the conclusion that $v_2$ is {\it
in-plane}.)

\subsection{The fourth harmonic \boldmath$v_4$\unboldmath}

Since elliptic flow $v_2$ is strong, the second order reaction plane $\Psi_2$
can be estimated with high precision at RHIC energies. This makes the study of
higher order flow feasible~\cite{v1v4Paper}.
\begin{figure}[h]
\vspace{0.5cm}
\resizebox{\FigFactor\textwidth}{!}{\includegraphics{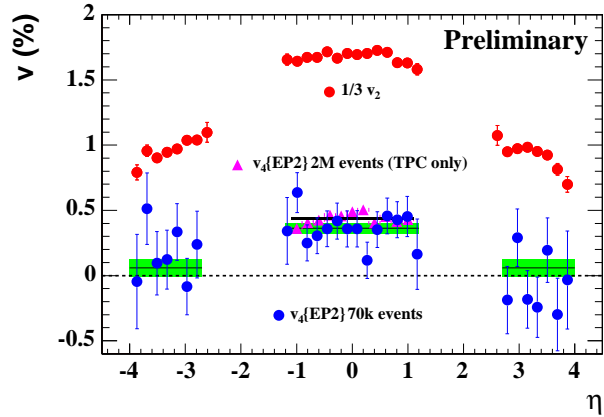}}
\caption{\label{v2v4}
Comparison of $v_2$ to $v_4\{\mathrm{EP}2\}$. The fourth harmonic (blue circles)
is consistent with zero at forward pseudorapidities while it reaches $(0.4 \pm
0.1)\%$ at mid-rapidities. Compared to the measurement of elliptic flow $v_2$
(red circles, scaled by a factor of 1/3 to fit on the plot) the fall-off from
$\eta = 0$ to $|\eta| = 3$ is stronger. The larger dataset available for the TPC
only (magenta triangles) confirms our measurement of $v_4\{\mathrm{EP}2\}$ at
mid-rapidities.}
\end{figure}

The fourth harmonic $v_4$ shows an average value of $(0.4\pm0.1)\%$ in
pseudorapidity coverage of the TPC ($|\eta| < 1.2$), see Fig.~\ref{v2v4}. In
contrast, its value of $(0.06\pm0.07)\%$ in the forward regions is consistent
with zero and we place a $2\sigma$ upper limit of 0.2\%. Therefore the fall-off
of $v_4$ from mid-rapidities to forward rapidities appears to be stronger than
for $v_2$. This behavior is consistent with scaling like $v_4\sim v_2^2$.

\section{Future developments}
First attempts to make use of the newly proposed method~\cite{LeeYangZeros}
utilizing Lee-Yang zeros are encouraging. This method eliminates higher order
non-flow contributions by construction. It is mathematically equivalent to the
\emph{all}-particle cumulant method $v\{\infty\}$ which takes into account all
higher order non-flow effects. The great advantage of the new method is its
simplicity and speed compared to the evaluation of the cumulants.

The upcoming RHIC run 4 will greatly enhance our data sample. With it we will
reduce our statistical uncertainties in the forward pseudorapiditiy region
significantly.

\acknowledgements{We thank the RHIC Operations Group and RCF at BNL, and the
NERSC Center at LBNL for their support. This work was supported in part by the
HENP Divisions of the Office of Science of the U.S.  DOE; the U.S. NSF; the BMBF
of Germany; IN2P3, RA, RPL, and EMN of France; EPSRC of the United Kingdom;
FAPESP of Brazil; the Russian Ministry of Science and Technology; the Ministry
of Education and the NNSFC of China; Grant Agency of the Czech Republic, FOM and
UU of the Netherlands, DAE, DST, and CSIR of the Government of India; the Swiss
NSF.}

\def\etal{\mbox{$\mathrm{\it et~al.}$}}

\end{document}